\documentclass[runningheads,orivec]{llncs}
\usepackage[T1]{fontenc}
\usepackage{graphicx}
\usepackage{color}
\usepackage[numbers,sort&compress]{natbib}
\usepackage{booktabs}
\usepackage{hyperref}
\usepackage{multirow}
\usepackage{pifont}
\usepackage{amsmath,amssymb,amsfonts}%
\usepackage{array}
\usepackage{tabularx}
\newcommand{\quotes}[1]{``#1''}
\newcommand{\xmark}{\ding{55}}%
\usepackage[nolist]{acronym}
\usepackage{enumitem}
\usepackage{subcaption}

\begin{document}
%
\title{Assessing Model-Agnostic XAI Methods against EU AI Act Explainability Requirements}
\titlerunning{Assessing XAI Methods against AI Act Requirements}
%
\author{Francesco Sovrano\inst{1}\orcidID{0000-0002-6285-1041} \and
Giulia Vilone\inst{2}\orcidID{0000-0002-4401-5664} \and
Michael Lognoul\inst{3}\orcidID{0009-0005-5137-8278}}
\authorrunning{F. Sovrano et al.}
%
\institute{University of Italian-Speaking Switzerland (USI)\\
\email{francesco.sovrano@usi.ch}\\ 
\and
Analog Devices International\\
\email{giulia.vilone@analog.com}
\and
CRIDS, NADI, University of Namur\\
\email{michael.lognoul@unamur.be}
}

\begin{acronym}
    \acro{AI}{Artificial Intelligence}
    \acro{XAI}{Explainable Artificial Intelligence}
    \acro{LEG}{Legal Explanatory Goal}
    \acro{MDR}{Medical Devices Regulation}
    \acro{ML}{Machine Learning}
    \acro{EU}{European Union}
    \acro{EDPB}{European Data Protection Board}
    \acro{GDPR}{General Data Protection Regulation}
    \acro{AIA}{Artificial Intelligence Act}
    \acro{DNN}{Deep Neural Network}
    \acro{DSA}{Digital Services Act}
    \acro{P2B}{Platform to Business Regulation}
    \acro{MiFID II}{Markets in Financial Instruments Directive II}
    \acro{SME}{small or medium-sized enterprise}
\end{acronym}
\maketitle              
\begin{abstract}
Explainable AI (XAI) has evolved in response to expectations and regulations, such as the EU AI Act, which introduces regulatory requirements on AI-powered systems. 
However, a persistent gap remains between existing XAI methods and society's legal requirements, leaving practitioners without clear guidance on how to approach compliance in the EU market.
To bridge this gap, we study model-agnostic XAI methods and relate their interpretability features to the requirements of the AI Act. We then propose a qualitative-to-quantitative scoring framework: qualitative expert assessments of XAI properties are aggregated into a regulation-specific compliance score. This helps practitioners identify when XAI solutions may support legal explanation requirements while highlighting technical issues that require further research and regulatory clarification.

\end{abstract}

\section{Introduction}

Opening up black-box models is increasingly driven by newly enacted regulatory frameworks, especially in the EU, that require the disclosure of their inferential process, implicitly requiring the use of \ac{XAI} for regulatory compliance \cite{bibal2021legal,sovrano2022survey,DBLP:conf/fat/PaniguttiHHLYJS23}. 
Yet, a \textit{transparency gap} \cite{DBLP:conf/ecai/GyevnarF023,bringas2025should}
has emerged between what XAI methods typically offer and what the law expects.
While \ac{XAI} is often defined narrowly (primarily as technical explanations of algorithms), the law treats explanations as tools for promoting accountability, human control over AI systems, and ensuring respect for human rights. 
Since this gap has largely remained unaddressed, AI practitioners and companies (especially the smaller ones) struggle to identify which XAI techniques best support compliance with EU legislation such as the AI Act \cite{europeancommission2024aiact}, a challenge with global implications due to the \quotes{Brussels Effect} \cite{bradford2020brussels}.

Existing surveys and research on \ac{XAI} have generally concentrated on algorithmic aspects \cite{arrieta2020explainable,vilone2021notions,holzinger2022information,nauta2023anecdotal,longo2024explainable}, and they lack a proposal for the systematic mapping of \ac{XAI} methods to legal requirements \cite{sovrano2022survey,DBLP:conf/cikm/CugnyACRT22,DBLP:conf/ecai/GyevnarF023,richmond2024explainable,fresz2024should}.
\cite{bringas2025should} connected XAI properties to legal explainability requirements, but their work lacks a systematic mapping of EU legislation to concrete XAI algorithms. \cite{sovrano2024aligning} proposed an alignment between the \ac{XAI} algorithms and the objectives pursued by laws that require AI explanations (i.e., legal explanatory goals), by considering the types of information (questions) that \ac{XAI} methods must provide (answer). 

This paper partially addresses this gap by analysing well-known, model-agnostic XAI methods, characterising their relevant properties, and classifying the information they extract. 
In parallel, drawing on the methodology of \cite{bringas2025should}, we examined the AI Act to identify its explanation obligations (and their explanatory goals). We then mapped these obligations to the capabilities of XAI methods. 
We operationalised this mapping by introducing a mixed-methods scoring framework for \ac{XAI} methods that combines qualitative assessments at the property level (e.g., faithfulness benchmarks and robustness analyses) with a quantitative aggregation procedure. This procedure weights each property according to the demands of the relevant legal requirement (mandatory, optional, or partial) and returns a regulation-specific \emph{compliance score}. 
The resulting framework, made up of the mapping and scoring processes, will enable practitioners to understand how mainstream interpretability techniques align with regulatory requirements and identify areas requiring further research.

\section{Research Methodology: Qualitative Assessment with Quantitative Aggregation} \label{sec:methodology} 

Before relating \ac{XAI} algorithms to legal explanatory requirements, it was necessary to clarify what is meant by an \textit{explanation}. The literature identifies at least five distinct definitions \citep{DBLP:conf/xai/SovranoV23}, each grounded in a different strand of contemporary philosophical theory.
In this work, we adopt the definition rooted in ordinary-language philosophy \cite{amsdottorato10943}, as it more closely aligns with how explanations are understood in legal contexts than the alternatives \citep{DBLP:conf/xai/SovranoV23,sovrano2022survey}. In this context, explanations are formalised as \quotes{answers to implicit or explicit questions that facilitate understanding of what is being explained}, a characterisation that is also consistent with established XAI literature \citep{liao2021question}.
Thus, an explanation should convey sufficient information to enable understanding by its intended audience. This perspective contrasts with approaches that require explanations to be tailored to an individual's mental model or restrict explanations to causal demonstrations. In legal settings, explanations are not necessarily required to be fully personalised \citep{wachter2017right}; for example, they need not be delivered in a recipient's native language when a different language is legally binding, and they may be addressed to law-defined categories such as the \quotes{average} patient or customer. Moreover, legal explanations may extend beyond purely causal accounts \citep{bibal2021legal,sovrano2020modelling}.

\begin{figure}
    \centering
    \includegraphics[width=\linewidth]{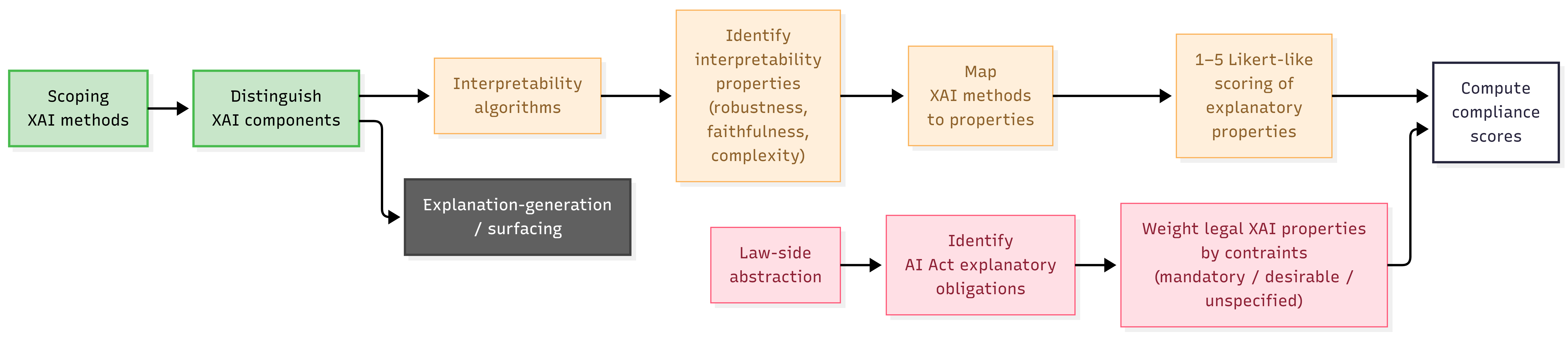}
    \caption{Methodology overview: starting from a set of XAI methods, the proposed framework extracts a set of explainability properties of these methods. Then, it matches these properties against the AI Act's legal explanatory requirements to rank XAI methods by their potential for legal compliance.}
    \label{fig:methodology}
\end{figure}

In line with the predominant literature, we split XAI approaches for black-box systems into two categories (see Figure~\ref{fig:methodology}): \textit{interpretability algorithms}, which extract explainable information from a black-box model, and \textit{explanation-generation} or \textit{surfacing} methods, which organise such information into the most suitable format for end users. This paper focuses only on interpretability algorithms because legal requirements typically concern the \emph{content} and a reasonable level of clarity, rather than prescribing specific surfacing procedures which are often context-dependent and vary across cultures, sectors, and organisations (according to internal procedures or documentation practices).

To map XAI methods to legal requirements, we first identified a set of interpretability properties from the literature as the hinge for this mapping (i.e., the law requires properties; XAI provides properties).
We relied on \cite{chen2022makes}, who cluster XAI interpretability properties into four main categories: \emph{robustness}, \emph{faithfulness}, \emph{complexity}, and \emph{homogeneity/fairness}.
We treated fairness (the last category) as part of faithfulness, which in this study means accurately representing the model's behaviour, even when that behaviour is unfair. So, if the model is biased or discriminatory toward certain groups, an explanation that correctly shows this is still faithful. 

We then identified the explanation obligations under the AI Act that apply to AI systems and, building on prior work \cite{sovrano2024aligning,bibal2021legal}, analysed the relevant \textit{provisions} (binding clauses) and \textit{recitals} (non-binding but interpretative).
This was done first through a \textit{literal reading} of transparency demands and objectives, then through a \textit{contextual reading} informed by regulators, case law, and scholarship. Using an inductive coding methodology \cite{fereday2006demonstrating}, a legal expert (co-author) grouped these into higher-level explanation requirements, which were coded along three dimensions: who is addressed, what is explained (individual output vs.\ system vs.\ global behaviour), and when (ex-ante/ex-post). Eventually, we translated these requirements into constraints on XAI properties (faithfulness, robustness, complexity), indicating for each whether it is mandatory, desirable, or unspecified.

Finally, we mapped each XAI method to the identified properties. Because these properties lack general, method-agnostic quantitative metrics, we adopted a mixed-methods assessment approach: we first performed a qualitative assessment of each algorithm on each property, and then quantitatively aggregated these assessments into a compliance score. For each algorithm, we rated the extent to which it exhibits each property on a 1-5 Likert-like scale (1 = not exhibited, 5 = fully exhibited). Following an inductive coding approach \cite{fereday2006demonstrating}, we reviewed the literature on each method to identify strengths, weaknesses, and existing evaluations, and assigned scores \emph{by comparison} (e.g., scoring SHAP above LIME, or RuleFit above decision trees, on faithfulness and completeness \cite{DBLP:conf/nips/LundbergL17,man2020best,friedman2008predictive,bramer2007avoiding,halabaku2024overfitting}). The resulting scores are therefore meaningful only relative to the set of XAI methods evaluated.

More specifically, the property scores for the analysed XAI methods were produced through a two-stage process. First, we prompted ChatGPT 4o (used through its web interface in its default configuration, with no hyper-parameters modified) to generate initial estimates according to the guidelines above. Second, the authors systematically reviewed and revised these initial scores. Any disagreements among the authors were resolved through joint re-examination of the relevant scientific literature. In the final version, all scores initially proposed by ChatGPT were modified by the authors.

Our scoring methodology is related to \emph{multiple-criteria decision analysis} (MCDA), a family of approaches for structuring decision problems involving multiple criteria and aggregating criterion-level assessments into an overall recommendation or ranking \cite{cinelli2020mcda,belton2002mcda}. The analogy is useful because our framework also evaluates a set of alternatives (XAI methods) against multiple dimensions (interpretability properties) and aggregates these assessments into a single score. However, our proposal is not a standard MCDA application. In classical MCDA, the criteria weights typically encode the preferences of a decision maker or stakeholder group, and the final output is a preference-sensitive recommendation among alternatives \cite{cinelli2020mcda,belton2002mcda}. By contrast, in our framework, the weights are derived from the normative relevance assigned by the AI Act to each property (e.g., mandatory, partial, or optional), and the resulting score is intended as a regulation-specific indicator of \emph{compliance potential}, not as a general utility score. For this reason, our method is better understood as an MCDA-inspired, legally grounded assessment framework.

After assigning the property scores, we estimate the potential for legal compliance of each XAI algorithm by quantitatively aggregating these scores with a mathematical formula (\S\ref{sec:xai_to_law_mapping:scoring_rule}) to calculate a \quotes{compliance score} in the range $[0,1]$, given a XAI method and an article of the AI Act.
Because the underlying property scores are qualitatively assigned, the resulting compliance scores presented in Section~\ref{sec:results} should not be interpreted as proof of legal compliance; they serve as valuable indicators for choosing the most promising XAI algorithms to generate legally compliant explanations.

\section{XAI Literature Review} \label{sec:methodology:xai_survey}
We examined survey papers to compile a comprehensive list of widely used post-hoc XAI methods for black-box models. We excluded inherently interpretable models used directly as predictors, but retained tree- and rule-based methods when they are used as surrogate explainers for black-box models, because in that setting they function as model-agnostic post-hoc explanation methods. 
We limited the search to studies published since 2022 with at least 10 citations (we waived this requirement for the latest work published within the previous 12 months) to capture the most recent and relevant trends in XAI, and we focused on methods that can be applied across model classes to enhance the generality of the proposed framework.
We searched Google Scholar using the terms:
\begin{quote}
\texttt{-site:arxiv.org} (\texttt{"survey" OR "systematic review"}) AND (\texttt{"explainable AI algorithms" OR "interpretable ML algorithms" OR "XAI algorithms"})\footnote{The command \quotes{-site:} excludes links from a specified URL, because of the \quotes{-} symbol.}
\end{quote}
The final pool included 30 published comprehensive surveys, collectively covering over 100 XAI algorithms.
Analysis revealed that approximately 82\% of these algorithms are cited in fewer than five surveys, and 91\% appear in fewer than ten. 
We focused on algorithms cited in at least 10 of the 30 surveys analysed.

In addition to the XAI methods, we had to extract a comprehensive set of explanation properties that may be required or highly desirable under the identified legal frameworks. 
We reviewed {30} peer-reviewed  \emph{survey papers on explanation properties} published since 2022. To retrieve them, we again performed a keyword-based search on Google Scholar:
\begin{quote} 
    \texttt{-site:arxiv.org} (\texttt{"survey" OR "systematic review"}) AND (\texttt{"explainable AI" OR "interpretable ML" OR "XAI"}) AND (\texttt{"explainability properties" OR "properties of XAI" OR "XAI properties" OR "properties of explanations" OR "explanation properties" OR "explanation criteria" OR "explanation characteristics"})
\end{quote}

\section{Background} \label{sec:background} 

\textbf{The AI Act}
\cite{europeancommission2024aiact} applies to AI systems marketed or used in the EU, regardless of where providers or deployers are located (Art.\ 2). \quotes{High-risk} systems are those covered by the sectoral legislation in Annex I and subject to third-party conformity assessment, as well as those listed in Annex III (e.g., credit scoring). 
They face three layers of explanation duties. First, providers must prepare detailed technical documentation covering design, logic, parameters, performance, and interpretability (Art.\ 11; Annex IV) \cite{sovrano2025simplifying}. 
Second, they must provide instructions for use, including system capabilities and limitations, intended purpose, guidance on interpreting outputs, and measures for human oversight (Arts.\ 13--14). 
Third, when Annex III high-risk systems support decision-making, deployers must provide affected persons \quotes{clear and meaningful explanations} of the AI's role and main decision elements, enabling them to exercise downstream rights (e.g., contesting the decision; Art.\ 86).

\textbf{Model-agnostic XAI methods} span several main families. Rule-based approaches, including decision trees, provide human-readable if-then logic but can be unstable and oversimplify complex decision boundaries~\cite{vilone2021classification}. 
Local surrogate methods, like LIME and SHAP, approximate the behaviour of a black-box around a given instance or via Shapley-value attributions, offering flexible explanations that are nevertheless sensitive to assumptions about local linearity, feature independence and the choice of neighbourhood or background distribution~\cite{DBLP:conf/kdd/Ribeiro0G16,DBLP:conf/nips/LundbergL17}. 
Global feature-effect tools such as PDP and ICE visualise how selected features influence model predictions: PDP provides a global, averaged view of feature effects across the dataset, whereas ICE disaggregates these effects into per-instance curves. Both typically rely on feature-independence assumptions and can struggle to capture higher-order interactions \cite{friedman2001greedy,goldstein2015peeking}.
Counterfactual methods, e.g., DiCE~\cite{DBLP:conf/fat/MothilalST20}, search for minimally changed inputs that flip a model's decision, yielding \quotes{what-if} explanations, but depend heavily on distance metrics and assumptions about which features can be modified independently. 

\textbf{Explainability Properties.}
Following \cite{chen2022makes}, we consider three main clusters of XAI properties, summarised in Table~\ref{tab:algorithm_property_cluster}. \emph{Faithfulness} (also referred to as \emph{fidelity} or \emph{correctness}) captures how accurately an explanation reflects the model's actual reasoning. It includes: 
(i) \emph{Necessity/No False Positives}, meaning that the explanation includes only those features or elements that genuinely influence the model's output; 
(ii) \emph{Sufficiency/No False Negatives}, meaning that it excludes only features or elements that are irrelevant to the output; 
and (iii) \emph{Completeness/Coverage}, meaning it covers the model's full rationale without omitting relevant factors.
\emph{Robustness} (also referred to as \emph{stability}) concerns the extent to which explanations are \emph{not overly sensitive} to small, irrelevant, or prediction-preserving changes to the input. It includes: 
(i) \emph{Stability/Continuity}, whereby similar instances or minor perturbations yield similar explanations; 
and (ii) \emph{Adversarial Robustness}, whereby explanations are not easily manipulated through adversarial examples.
Finally, \emph{Complexity} (a.k.a.\ Compactness, Minimality) concerns the complexity of an explanation itself, including \emph{Sparsity/Few Features} and \emph{Rule/Predicate Size}, as well as its \emph{Dimensionality/Granularity/Level of Detail}, i.e., whether the explanation is aggregated or highly specific.

\begin{table}
    \scriptsize
    \caption{\footnotesize Main XAI interpretability properties from \cite{chen2022makes}.} \label{tab:algorithm_property_cluster}
    \centering
    \begin{tabular}{m{1.9cm}|m{2.4cm}|m{7.5cm}}
        \toprule
        Cluster & Description & Sub-properties \\
        \midrule
        \textbf{Faithfulness/ Correctness} & How accurately an explanation reflects the underlying model's true reasoning. & 
            \textbf{Necessity/No False Positives:} Only features or elements that actually influence the model's outcome. \newline
             \textbf{Sufficiency/No False Negatives:} Only features irrelevant to the model's outcome are ignored. \newline
             \textbf{Completeness/Coverage:} The explanation covers the model's full rationale, not leaving out relevant factors.\\ \hline
        \textbf{Robustness/ Sensitivity} & Stability of the explanation under changes in inputs. &
            \textbf{Stability/Continuity:} Similar instances or minor perturbations yield similar explanations.\newline
            \textbf{Adversarial Robustness:} Explanation should not be easily manipulated by adversarial examples. \\
            \hline
        \textbf{Complexity} & Complexity of an explanation. & 
            \textbf{Sparsity/Few Features} \& \textbf{Rule/Predicate Size}. \newline
            \textbf{Dimensionality/Granularity/Level of Detail}.\\
        \bottomrule
    \end{tabular}
\end{table}

\section{Property-to-Algorithm Mapping} 
Following the methodology in \S\ref{sec:methodology}, we scored 
widely used model-agnostic XAI algorithms on a 1--5 Likert scale
against the faithfulness, robustness, and complexity sub-properties identified in \S\ref{sec:background}, where 1 means not exhibited and 5 means fully exhibited. 

\textit{Decision trees} \cite{bastani2017interpreting} 
offer fine-grained \quotes{level of detail} (5), but are highly sensitive to perturbations (\quotes{stability} 1); unpruned trees overfit, increasing \quotes{false positives} and reducing \quotes{adversarial robustness} (both 2). 
\textit{RuleFit} \cite{friedman2008predictive}, combining multiple decision-trees, improves \quotes{completeness} and \quotes{level of detail} (4) over single trees. \quotes{Sparsity} drops to 2, but it is more \quotes{stable} (3 vs 1) than Decision Trees, while compared to SHAP it trades some \quotes{faithfulness} (3 vs 5) for greater rule-based interpretability.
\textit{RuleSHAP} \cite{sovrano2025can} uses SHAP-guided weights to make RuleFit more faithful, raising \quotes{no false positives} and \quotes{no false negatives} to 4; it improves on RuleFit while remaining slightly less faithful than SHAP (4 vs 5).

\textit{PDPs} \cite{friedman2001greedy} average predictions over data to give smooth global curves with high \quotes{stability} (4). Their main drawbacks are visual clutter (many plots needed; \quotes{sparsity} 2). 
They provide less instance-level recall than \textit{ICE} \cite{goldstein2015peeking}, which disaggregates PDPs into per-instance curves on the same grid \cite{friedman2001greedy}, revealing heterogeneity and interactions and increasing its \quotes{no false negatives} score to 4. ICE shares \quotes{level of detail} (4) with PDPs but suffers from overplotting (\quotes{sparsity} and \quotes{completeness} 2) and lower \quotes{stability}, but it is more stable than LIME.

\textit{LIME} \cite{DBLP:conf/kdd/Ribeiro0G16} produces moderately sparse local linear surrogates but no dimension above 3. Randomized perturbations cause high variance (\quotes{stability} 1), and explanations are attackable (\quotes{adversarial robustness} 1). 
It is less accurate than SHAP (\quotes{no false positives/negatives} 2 vs 5) \cite{man2020best}; Anchors improves its precision \cite{DBLP:conf/aaai/Ribeiro0G18}.
\textit{SHAP} \cite{DBLP:conf/nips/LundbergL17} achieves the top scores in \quotes{no false positives/negatives} (both 5) for exact values (approximate schemes, e.g., KernelSHAP, can reduce faithfulness), while satisfying the standard additive axioms of local accuracy and missingness. Its remaining dimensions are all $\geq 3$; compared to LIME, it offers much stronger faithfulness while remaining less rule-like (\quotes{level of detail} 3) than Decision Trees \cite{bastani2017interpreting}.

\textit{Anchors} \cite{DBLP:conf/aaai/Ribeiro0G18} learns short if-then rules with high-precision guarantees via randomised perturbations and bandit search, targeting high \quotes{sparsity} (5) and \quotes{no false positives} (4) \cite{DBLP:conf/aaai/Ribeiro0G18}. The stochastic search yields very low \quotes{stability} (1); relative to CEM, it matches \quotes{sparsity} but offers weaker \quotes{adversarial robustness} (3 vs 4). 
\textit{CEM}  \cite{DBLP:conf/nips/DhurandharCLTTS18} returns pertinent positives/negatives that are minimally sufficient/necessary, giving \quotes{no false positives} 5, high \quotes{sparsity}, and strong \quotes{adversarial robustness} (4) via $\ell_1$/$\ell_2$ penalties and manifold regularisation. 
Its two non-convex optimisation problems with several hyperparameters lead to low \quotes{stability} (1), but compared to Anchors it attains higher \quotes{completeness} and slightly better robustness.
\textit{DiCE} \cite{DBLP:conf/fat/MothilalST20} enforces validity through a margin constraint in a single diversity-aware optimisation, achieving strong \quotes{no false positives} (5), good \quotes{sparsity} (4), and high \quotes{adversarial robustness} (4) when constraints are feasible. As with CEM, non-convexity, random initialisation, and many iterations yield very low \quotes{stability} scores.

\begin{table}
    \centering
    \footnotesize
    \caption{
    Exact scores (1--5) for the faithfulness, robustness, and complexity sub-properties of Table~\ref{tab:algorithm_property_cluster}.}
    \resizebox{\linewidth}{!}{
        \begin{tabular}{l|ccc|cc|cc} 
        \toprule
        \multirow{2}{*}{\textbf{XAI Algorithm}} & \multicolumn{3}{c|}{\textbf{Faithfulness}} & \multicolumn{2}{c|}{\textbf{Robustness}} & \multicolumn{2}{c}{\textbf{Complexity}} \\
         & \textbf{No FP} & \textbf{No FN} & \textbf{Completeness} & \textbf{Stability} & \textbf{Adv. Rob.} & \textbf{Sparsity} & \textbf{Level of Detail} \\ 
        \midrule
        Decision Trees & 2 & 3 & 3 & 1 & 2 & 3 & 5 \\
        RuleFit & 3 & 3 & 4 & 3 & 3 & 2 & 4 \\
        RuleSHAP & 4 & 4 & 4 & 3 & 3 & 3 & 4 \\
        PDP & 3 & 3 & 3 & 4 & 3 & 2 & 4 \\
        ICE & 3 & 4 & 2 & 3 & 3 & 2 & 4 \\
        LIME & 2 & 2 & 2 & 1 & 1 & 3 & 2 \\
        SHAP & 5 & 5 & 3 & 4 & 4 & 3 & 3 \\
        Anchors & 4 & 3 & 3 & 1 & 3 & 5 & 3 \\
        CEM & 5 & 3 & 4 & 1 & 4 & 4 & 3 \\
        DiCE & 5 & 3 & 3 & 1 & 4 & 4 & 3 \\
        \bottomrule
        \end{tabular}
    }
    \label{tab:property_xai_scores_model_agnostic}
\end{table}

\begin{table}
    \centering
    \footnotesize
    \caption{
    Scores (1--5) and brief justifications for model-agnostic \ac{XAI} methods; for each algorithm, we highlight up to three highest- and lowest-scoring sub-properties (\quotes{Best at} / \quotes{Worst at}), consistently with Table~\ref{tab:property_xai_scores_model_agnostic}.}
    \resizebox{\linewidth}{!}{
        \begin{tabular}{l|l|l}
            \toprule
            \begin{tabular}[c]{@{}l@{}}\textbf{XAI }\\\textbf{Algorithm}\end{tabular} & \textbf{Best at (score \textgreater{} 3)} & \textbf{Worst at (score \textless{} 3)}\\
            \midrule
            \begin{tabular}[c]{@{}l@{}}Decision\\ Trees\end{tabular}
            & \begin{tabular}[c]{@{}l@{}}\textbf{Level of Detail (5):} Very fine-grained, per-leaf\\ explanations.\end{tabular}
            & \begin{tabular}[c]{@{}l@{}}\textbf{Stability (1):} Highly sensitive to small\\ data changes.\\
            \textbf{No FP (2):} Splits may reflect spurious local\\ patterns.\\
            \textbf{Adv. Rob. (2):} Vulnerable to adversarial\\ perturbations.\end{tabular} \\
            \midrule
            RuleFit
            & \begin{tabular}[c]{@{}l@{}}\textbf{Completeness (4):} Blends rules and linear terms\\ for broad coverage.\\
            \textbf{Level of Detail (4):} Detailed rule+coefficient\\ breakdown.\end{tabular}
            & \begin{tabular}[c]{@{}l@{}}\textbf{Sparsity (2):} Hundreds of trees still yield\\ many rules.\end{tabular} \\
            \midrule
            RuleSHAP
            & \begin{tabular}[c]{@{}l@{}}\textbf{No FP (4):} SHAP-driven rules are highly faithful.\\
            \textbf{No FN (4):} Captures most true influences.\\
            \textbf{Completeness (4):} Rules cover the main decision\\ rationale well.\end{tabular}
            & \textit{(none \textless{} 3)} \\
            \midrule
            PDP
            & \begin{tabular}[c]{@{}l@{}}\textbf{Stability (4):} Averaging yields smooth,\\ repeatable curves.\\
            \textbf{Level of Detail (4):} Full feature-effect curves.\end{tabular}
            & \begin{tabular}[c]{@{}l@{}}\textbf{Sparsity (2):} Presents every feature's curve.\end{tabular} \\
            \midrule
            ICE
            & \begin{tabular}[c]{@{}l@{}}\textbf{No FN (4):} Reveals more individual effects.\\
            \textbf{Level of Detail (4):} Full per-instance effect\\ curves.\end{tabular}
            & \begin{tabular}[c]{@{}l@{}}\textbf{Completeness (2):} One feature at a time;\\ limited coverage.\\
            \textbf{Sparsity (2):} Still shows a full curve for each\\ data point.\end{tabular} \\
            \midrule
            LIME
            & \textit{(none \textgreater{} 3)}
            & \begin{tabular}[c]{@{}l@{}}\textbf{Stability (1):} Extremely sensitive to sampling.\\
            \textbf{Adv. Rob. (1):} Easily manipulated.\\
            \textbf{Completeness (2):} Local surrogate covers only a\\ narrow neighborhood.\end{tabular} \\
            \midrule
            SHAP
            & \begin{tabular}[c]{@{}l@{}}\textbf{No FP (5):} Only truly contributive features.\\
            \textbf{No FN (5):} Captures all positive/negative\\ contributions.\\
            \textbf{Stability (4):} Additive attributions are usually\\ stable for a fixed model and input.\end{tabular}
            & \textit{(none \textless{} 3)} \\
            \midrule
            Anchors
            & \begin{tabular}[c]{@{}l@{}}\textbf{Sparsity (5):} Very compact anchors.\\
            \textbf{No FP (4):} High precision by design.\end{tabular}
            & \begin{tabular}[c]{@{}l@{}}\textbf{Stability (1):} Highly stochastic solver.\end{tabular} \\
            \midrule
            CEM
            & \begin{tabular}[c]{@{}l@{}}\textbf{No FP (5):} Produces highly faithful minimal\\ changes.\\
            \textbf{Completeness (4):} Provides richer rationale than\\ basic counterfactuals.\\
            \textbf{Adv. Rob. (4):} Explanations remain valid under\\ perturbations.\end{tabular}
            & \begin{tabular}[c]{@{}l@{}}\textbf{Stability (1):} Outputs vary with solver seed.\end{tabular} \\
            \midrule
            DiCE
            & \begin{tabular}[c]{@{}l@{}}\textbf{No FP (5):} Correct what-if explanations.\\
            \textbf{Adv. Rob. (4):} Valid under adversarial\\ changes.\\
            \textbf{Sparsity (4):} Optimizes minimal feature\\ changes.\end{tabular}
            & \begin{tabular}[c]{@{}l@{}}\textbf{Stability (1):} Counterfactuals change\\ across runs.\end{tabular} \\
            \bottomrule
        \end{tabular}
    }
    \label{tab:property_xai_map_model_agnostic}
\end{table}


\section{Legal Compliance Assessment of XAI Methods} \label{sec:xai_to_law_mapping} \label{sec:xai_to_law_mapping:scoring_rule}

\begin{table}
  \scriptsize
  \centering
  \caption{\footnotesize Explainability property requirements demanded by the AI Act. Symbols: $\checkmark$ = required; $\checkmark$ (preferable) = optional but recommended; $\checkmark$ (partial) = partially required; \xmark = not required.}
  \label{tab:property-mapping}
  \resizebox{\linewidth}{!}{
      \begin{tabular}{ll|ccc}
        \toprule
        \multicolumn{2}{l|}{} & Art.\ 86 & Arts.\ 13--14  & Art.\ 11 \& Annex IV  \\ 
        \hline
        \multirow{3}{*}{\textbf{Faithfulness}}                                          & No
        false positives   & \checkmark  & preferable & \checkmark  \\
        & No false negatives     & \checkmark & \checkmark  & \checkmark \\
        & Completeness & \xmark   &\checkmark (reasonable)  & \checkmark \\ 
        \hline
        \multirow{2}{*}{\textbf{Robustness}}                              & Stability & \checkmark    & \checkmark    & \checkmark  \\
        & Adversarial robust.   & \checkmark (partial)   & \checkmark  & \checkmark \\
        
        \hline
        \multirow{2}{*}{\textbf{Complexity}}                               & Sparsity   & \checkmark & \xmark    & \xmark \\
        & Detailed    & \xmark  & \xmark   & \checkmark  \\ 
        \hline
        \multirow{3}{*}{\begin{tabular}[c]{@{}l@{}}\textbf{Other Prop.}\end{tabular}}
          & Scope
          & local  & both & global    \\
          & Stage
          & ex-post   & both  & ex-ante  \\
          \bottomrule
      \end{tabular}
  }
\end{table}

Table~\ref{tab:property-mapping} translates explanation duties demanded by the AI Act into a harmonised set of interpretability properties. This mapping is based on a close reading of the legal text, its objectives, and its existing interpretations (where applicable). The table can serve as a compliance checklist and a negotiation tool between legal and technical teams. Once the minimum thresholds for each property are frozen, engineers can select or tailor the most suitable XAI methods to meet or exceed the legal minimums.

We operationalise the high-level mapping in Table~\ref{tab:property-mapping} through the quantitative component of our mixed-methods approach, by applying a scoring procedure to identify, for each regulatory column, the XAI algorithm(s) that best satisfy the required interpretability properties, as described below.  
To do so, we combine three information sources:
\begin{enumerate}[label=\textbf{D\arabic*}]
  \item\label{D1} The \emph{legal demand side} (properties that carry a $\checkmark$ or equivalent wording in Table~\ref{tab:property-mapping});
  \item\label{D2} The \emph{algorithmic supply side} (Table~\ref{tab:property_xai_scores_model_agnostic}, which reports the full score matrix for the faithfulness, robustness, and complexity sub-properties; Table~\ref{tab:property_xai_map_model_agnostic} provides a compact qualitative summary);
  \item\label{D3} The \emph{procedural fit} (does the method work at the correct \emph{scope}, local or global, and \emph{stage}, ex-ante or ex-post?).
\end{enumerate}


For a legislation article $r$, let's define:
\begin{itemize}
    \item $\mathcal{R}_r=\{p_1,\dots,p_{m_r}\}$ be the set of the $m_r$ interpretability-property \emph{categories} $p_i$ (\textit{faithfulness}, \textit{robustness}, \textit{complexity}) that the legal text marks as required;
    \item $\mathcal{S}_{p_i}^r=\{s_{i1},\dots,s_{iK_{p_i,r}}\}$ the set of sub-properties $s_{ij}$ that the legislation article $r$ \emph{requires} of $p_i$. 
\end{itemize}
Each sub-property has a legal strength factor:
\begin{equation}
\lambda_s
  \;=\;
  \begin{cases}
    1      &\text{mandatory (\checkmark)},\\[2pt]
    0.75   &\text{optional/preferably},\\[2pt]
    0.5    &\text{partial}.
  \end{cases}
\end{equation}

$\operatorname{Score}(a,s)$ is the \quotes{interpretability} score achieved by an XAI algorithm $a$ for the explanatory property $s$ (see Table~\ref{tab:property_xai_scores_model_agnostic}), normalised in $[0,1]$. The weight
$w_{p_i}(a,r) \in [0,1]$, representing the \emph{average fraction of the (weighted) sub-properties of $p_i$ that $a$ satisfies}, is computed as:
\begin{equation}
w_{p_i}(a,r)
 = 
\frac{\displaystyle\sum_{s\in\mathcal{S}_{p_i}^r}
\lambda_s \cdot \operatorname{Score}(a,s)}
{\displaystyle\sum_{s\in\mathcal{S}_{p_i}^r}\lambda_s}.
\label{eq:wp}
\end{equation}
If an algorithm does not advertise any strength in a required sub-property, that dimension simply contributes $0$ to the numerator.
With this refined weight, the overall \emph{legislation-specific compliance score} becomes

\begin{equation}
S(a,r) = 
\frac{1}{|\mathcal{R}_r|}
\sum_{p_i\in\mathcal{R}_r}
w_{p_i}(a,r)
\cdot \mathbf{1}\!\Bigl[\text{\ref{D3} satisfied}\Bigr],
\label{eq:score}
\end{equation}

We assign the same priority or weight to each category, treating all properties as equal, even if some appear more frequently in the literature or in the law. However, this can be modified in the formula by using a weighted average.

The filtering step \ref{D3} is enforced \emph{before} scoring: an XAI algorithm contributes to the compliance score in equation \eqref{eq:score} only if its \emph{scope} and \emph{stage} intersect with the corresponding descriptors of article $r$. Formally, let $\mathrm{scope}(a)$ and $\mathrm{stage}(a)$ be, respectively, the scope and stage of algorithm $a$, and analogously $\mathrm{scope}(r)$ and $\mathrm{stage}(r)$ for the article. Then,
\begin{equation}
\begin{aligned}
\mathbf{1}[\text{\ref{D3} satisfied}]
  &= \mathbf{1}\Bigl[(\mathrm{scope}(a) \cap \mathrm{scope}(r)) \neq \varnothing \\
  &\qquad\land\; (\mathrm{stage}(a) \cap \mathrm{stage}(r)) \neq \varnothing
     \Bigr].
\end{aligned}
\end{equation}

To make criterion \ref{D3} reproducible, Table~\ref{tab:d3-profiles} assigns each XAI method an explicit \emph{scope} and \emph{stage} profile. We classify \emph{scope} according to the method's native explanatory unit. We classify \emph{stage} functionally: a method is \emph{ex-post} when it requires a realised case or prediction to generate an explanation, \emph{ex-ante} when it can characterise the model without reference to a specific executed input--output pair, and \emph{both} when it natively supports both uses.
Under this convention, LIME, Anchors, CEM, and DiCE are local and ex-post because they explain a specific prediction or input instance. PDP is global and ex-ante because it summarises the expected model response as a function of one or more features over the data distribution. SHAP supports both local attributions and global summaries. ICE is treated as local/global and both because it can be used either to inspect a specific case or to visualise heterogeneity across many cases. Surrogate decision trees and rule-based surrogate methods are treated here as local/global and both, because they can be used both to document model behaviour globally and to explain a specific case through rule paths or local rule contributions.
Accordingly, criterion \ref{D3} is a procedural admissibility filter: methods whose native scope or stage does not match the legal provision are excluded before scoring. In particular, local ex-post methods are not admissible for provisions that require global ex-ante documentation, while global ex-ante methods are not admissible for provisions that require local ex-post justifications.

\begin{table}[t]
\footnotesize
\centering
\caption{Procedural-fit descriptors used for criterion \ref{D3}. Scope is classified by the native explanatory unit of each method. Stage is classified functionally: \emph{ex-post} when the method requires a realised case or prediction to generate the explanation; \emph{ex-ante} when it can characterise the model without reference to a specific executed input--output pair; \emph{both} when the method natively supports both uses.}
\label{tab:d3-profiles}
\begin{tabular}{@{}lcc@{}}
\toprule
\textbf{XAI method} & \textbf{Scope} & \textbf{Stage} \\
\midrule
Decision Trees (surrogate) & local, global & both \\
RuleFit & local, global & both \\
RuleSHAP & local, global & both \\
PDP & global & ex-ante \\
ICE & local, global & both \\
LIME & local & ex-post \\
SHAP & local, global & both \\
Anchors & local & ex-post \\
CEM & local & ex-post \\
DiCE & local & ex-post \\
\bottomrule
\end{tabular}
\end{table}

Equation~\eqref{eq:score} guarantees the following properties:
\begin{enumerate}[label=\textbf{P\arabic*}]
    \item \textbf{Legal strength is respected}: each sub-property is weighted by its statutory factor~$\lambda_s$, so failing a \emph{mandatory} requirement reduces the weight strictly more than failing an \emph{optional/partial} one.
    \item \textbf{Normalisation removes cardinality bias}: division by $\sum_{s\in\mathcal{S}_{p_i}^r}\lambda_s$ ensures that categories containing more sub-properties cannot dominate purely by size; two algorithms satisfying the same \emph{fraction} of \emph{mandatory} and \emph{optional} requirements will receive the same $w_{p_i}(a,r)$  even when $\lvert\mathcal{S}_{p_i}^r\rvert$ differs.
    \item \textbf{Partial compliance is credited}: $\operatorname{Score}(a,s) \in [0,1]$ allows nuanced assessment, avoiding the all-or-nothing artefacts of Boolean checklists.
    \item \textbf{Zero reward for silence}: if the documentation of the XAI algorithm $a$ is silent about a required sub-property $s$, the term $\operatorname{Score}(a,s)=0$ yields no contribution, discouraging strategic omission.
    \item \textbf{The score has bounded interpretability}: by construction, $S(a,r) \in [0,1]$ as $w_{p_i}(a,r) \in [0,1]$, so the final score can be interpreted as a percentage of regulatory alignment.
\end{enumerate}

Taken together, these properties make the compliance score of equation \eqref{eq:score} \emph{comparable across algorithms, transparent to auditors, and sensitive to the legal priorities encoded in the legal strength factor~$\lambda_s$}.

\section{Results \& Discussion}\label{sec:results}
We conducted a systematic legal analysis of the provisions and recitals of the AI Act (Art.\ 11, Annex IV, Arts.\ 13--14, and Art.\ 86) that explicitly or implicitly require explanations about algorithmic systems. 
Table \ref{tab:property-mapping} links each legal provision to mandatory or desirable XAI properties. When laws are silent, properties are marked \quotes{not specified}; conditional language is captured accordingly. Across all statutes, \textit{faithfulness} and \textit{robustness} form the non-negotiable baseline; \textit{complexity} limits depend on the audience.

Using equation \eqref{eq:score}, we computed $S(a,r)$ for all algorithms in Table~\ref{tab:property_xai_scores_model_agnostic}, applying the procedural-fit filter of criterion \ref{D3} using the scope/stage profiles reported in Table~\ref{tab:d3-profiles}.
Table~\ref{tab:legal-top3} maps property requirements to the best-scoring model-agnostic XAI tools according to their compliance scores, rescaled from 1--5 to the $[0,1]$ range. Rather than a single best method, these techniques span a structured design space. 
Tree- and rule-based explainers (\textit{surrogate decision trees}, \textit{RuleFit}, \textit{RuleSHAP}, \textit{Anchors}) prioritise human-readable, often sparse decision paths or rule sets when \quotes{sparsity} and \quotes{level of detail} dominate. Attribution- and perturbation-based methods (\textit{LIME}, \textit{SHAP}) maximise \quotes{faithfulness} under clear axioms but differ in \quotes{robustness} and computational cost, with SHAP more reliable but slower.

Across provisions, Table~\ref{tab:legal-top3} shows a clear pattern: methods with strong \textit{faithfulness} and \textit{robustness} repeatedly rank near the top across the AI Act's explanatory requirements. 
For Art.~86 (local, ex-post justifications), SHAP, RuleSHAP, and CEM support highly faithful instance-level explanations; here, counterfactual examples, trees, and rule-based surrogates can be layered on top as \emph{presentation} devices when audiences require low \textit{complexity}, without using global methods as the primary compliance backbone.
For Arts.~13--14 (mixed local/global, ongoing information duties), SHAP and RuleSHAP again dominate on \textit{faithfulness}, while PDP provides robust, easily documentable global trends; in practice, providers and deployers can front-load simpler rule lists or decision trees in user-facing interfaces and keep SHAP-based diagnostics in the technical file, ensuring that reductions in \textit{sparsity} or \textit{level of detail} are an explicit design choice rather than an undocumented side effect. The explicit D3 profiles of Table~\ref{tab:d3-profiles} also clarify why PDP disappears from Art.~86 and why local ex-post methods such as Anchors, CEM, and DiCE disappear from Art.~11 and Annex~IV: the former is treated as global/ex-ante, whereas the latter are treated as local/ex-post.

Art.~11 and Annex~IV (global, ex-ante documentation) and Art.~86 (local, ex-post justifications) additionally constrain \textit{complexity}, but they do so over different admissible sets. For Art.~11 and Annex~IV, local ex-post methods such as Anchors, CEM, and DiCE are excluded by criterion \ref{D3}, so the best admissible low-complexity candidates are Decision Trees, RuleFit, and RuleSHAP. A sensible pattern is therefore to use SHAP or RuleSHAP as the primary evidential tool for risk management and performance monitoring, and then compress these explanations into tree- or rule-based summaries for Annex~IV-style documentation.
Overall, the rankings operationalise legal requirements as engineering choices but do not, by themselves, guarantee compliance: scores are context-dependent and assume careful implementation (e.g., avoiding degenerate KernelSHAP settings, calibrating counterfactual constraints), and they must be integrated with broader governance measures, such as data governance, human oversight, and documentation practices.

\begin{table}
  \footnotesize
  \centering
  \caption{Best XAI methods per provision and evaluation property (higher is better).}
  \label{tab:legal-top3}
  \begin{tabular}{llp{7.5cm}}
    \toprule
Provision & Property & Top-3 algorithms (score) \\ 
\midrule
\multirow{3}{*}{Art.~86} & Robustness & SHAP (0.80), RuleFit (0.6) \\
 & Faithfulness & SHAP (1.), RuleSHAP (0.80), CEM (0.80) \\ 
 & Complexity & Anchors (1.), CEM (0.80), DiCE (0.80) \\ 
\cline{2-3}
 & All & SHAP (0.8), Anchors (0.68), RuleSHAP (0.67) \\ 
\midrule
\multirow{3}{*}{Arts.~13-14} & Robustness & SHAP (0.80), PDP (0.70), RuleFit (0.60) \\
 & Faithfulness & SHAP (0.88), RuleSHAP (0.80), CEM (0.78) \\ 
\cline{2-3}
 & All & SHAP (0.84), RuleSHAP (0.70), PDP (0.65) \\ 
\midrule
\multirow{4}{*}{\begin{tabular}[c]{@{}l@{}}Art.~11 \\Annex IV\end{tabular}} & Robustness & SHAP (0.80), PDP (0.70), RuleFit (0.60) \\
 & Faithfulness & SHAP (0.87), RuleSHAP (0.8), RuleFit (0.67) \\
 & Complexity & Decision Trees (1.), RuleFit (0.8),  RuleSHAP (0.8) \\ 
\cline{2-3}
 & All & SHAP (0.76), RuleSHAP (0.73), PDP (0.70) \\
\bottomrule
  \end{tabular}
\end{table}

Lastly, we assessed the sensitivity of the proposed compliance scores to changes in the legal strength factors $\lambda$ by applying a correction factor $\delta$ that varied between $-0.2$ and $0.2$, representing a $20\%$ maximum variation in the legal strength factors. The corrected $\lambda$ values were limited within the range $[0,1]$ and computed as:
\begin{equation}
    \tilde{\lambda_s} = min(max(\lambda_s + \delta, 0.0), 1.0)
\end{equation}

The variation of the $\lambda$ factors did not have any impact on the final robustness and faithfulness compliance scores related to Art.\ 11 \& Annex IV of the AI Act. This is due to the fact that this article does not have partial requirements on these two XAI properties (see Table~\ref{tab:property-mapping}): both properties are fully mandatory ($\lambda = 1$ for every sub-property), so all scores are shifted up or down by the same $\delta$ variation which is, subsequently, cancelled out by the normalisation process.
The same phenomenon occurs for robustness and complexity when related to the Arts.\ 13--14. 

\begin{figure}
    \centering
    \begin{subfigure}[b]{0.49\textwidth}
    \includegraphics[width=\linewidth]{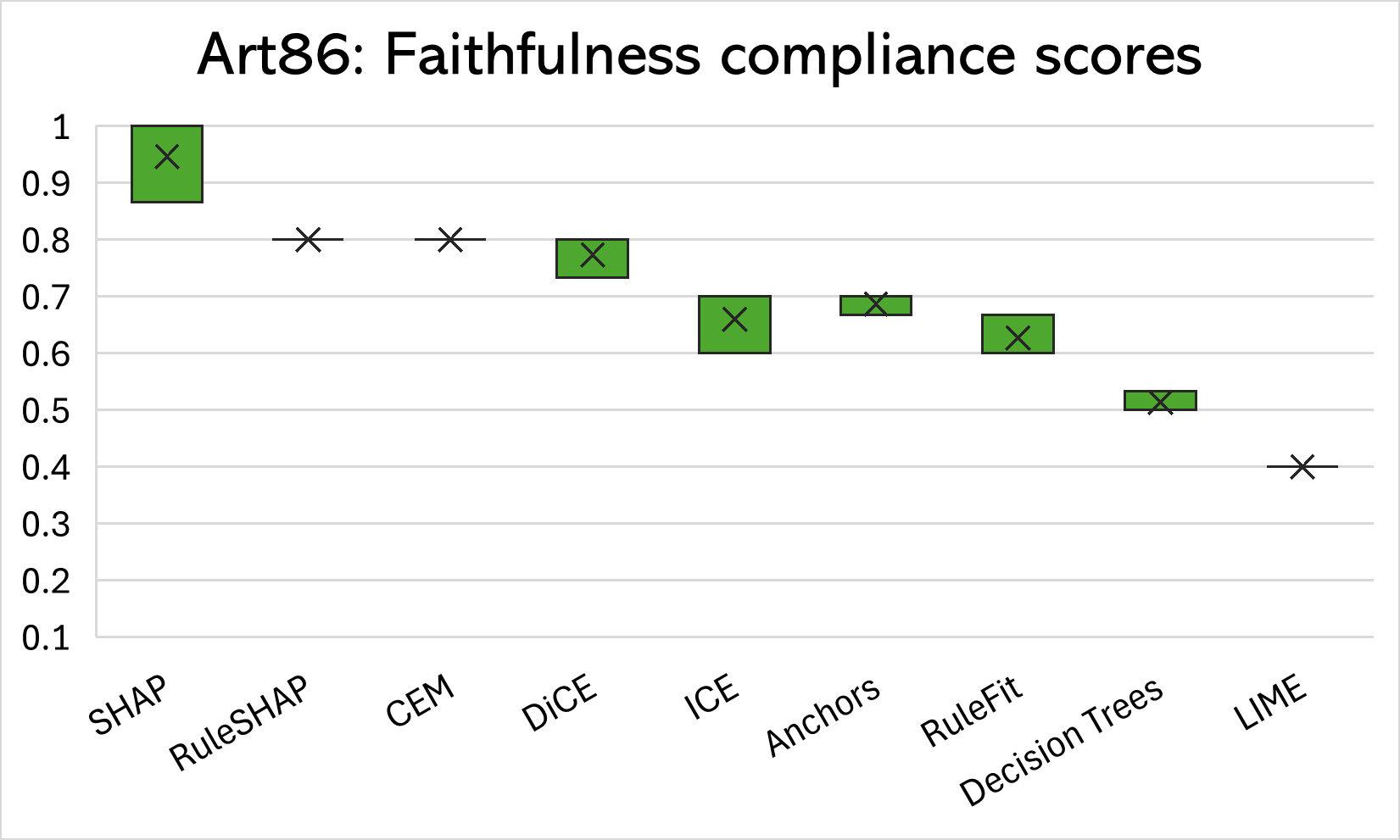}
    \caption{}
    \end{subfigure}
    \begin{subfigure}[b]{0.4\textwidth}
    \includegraphics[width=\linewidth]{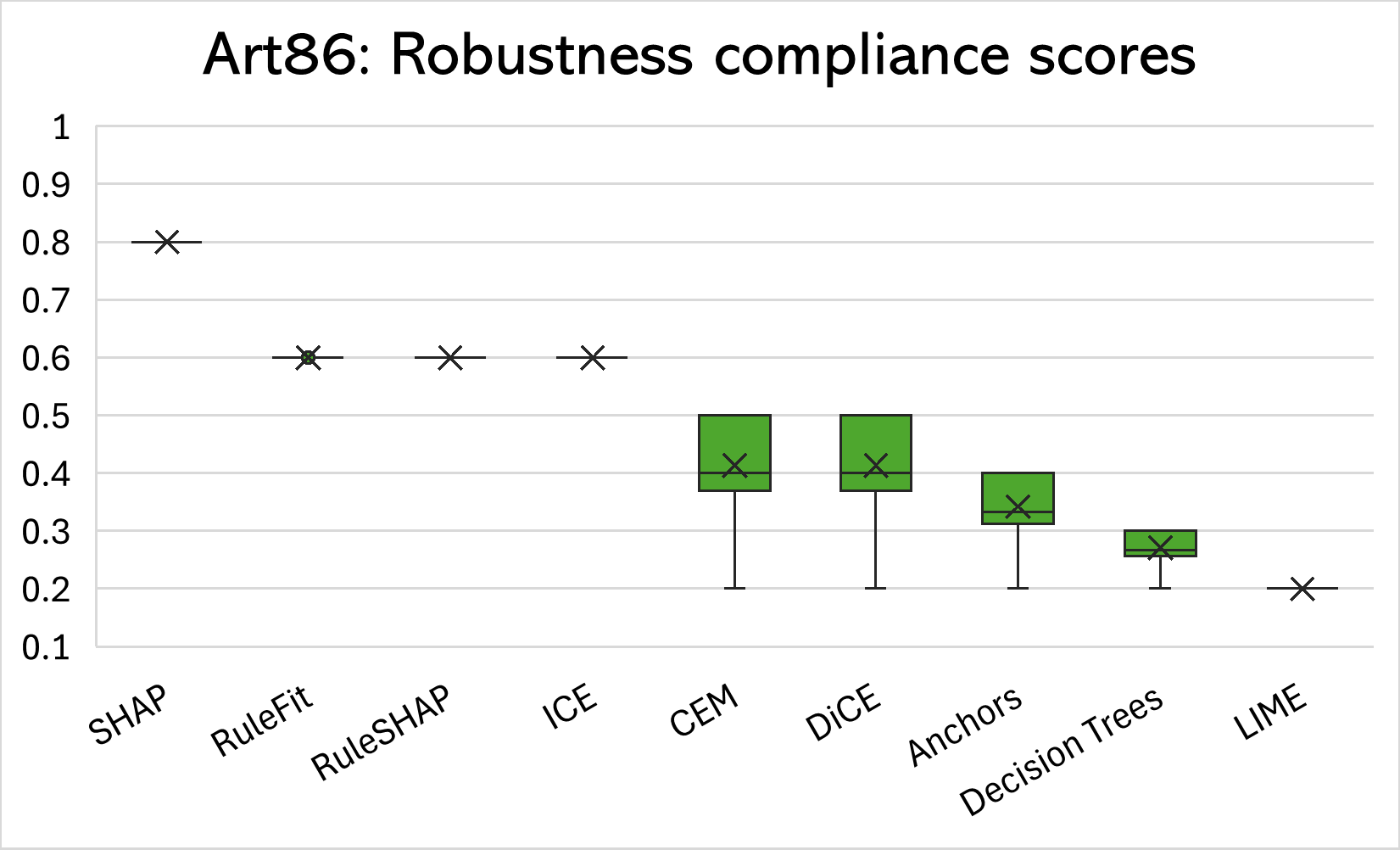}
    \caption{}
    \end{subfigure}
    \begin{subfigure}[b]{0.4\textwidth}
    \includegraphics[width=\linewidth]{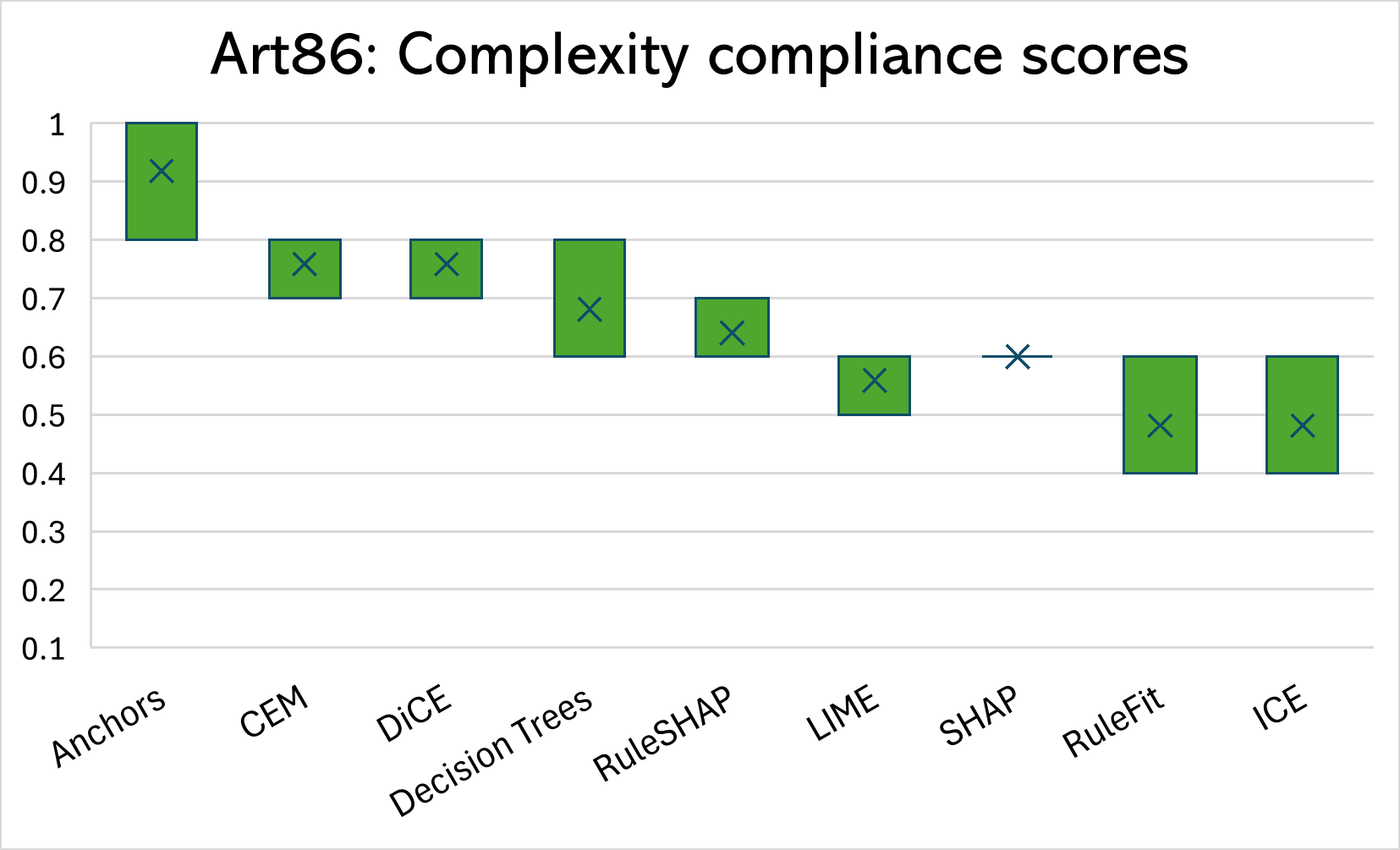}
    \caption{}
    \end{subfigure}
    \begin{subfigure}[b]{0.49\textwidth}
    \includegraphics[width=\linewidth]{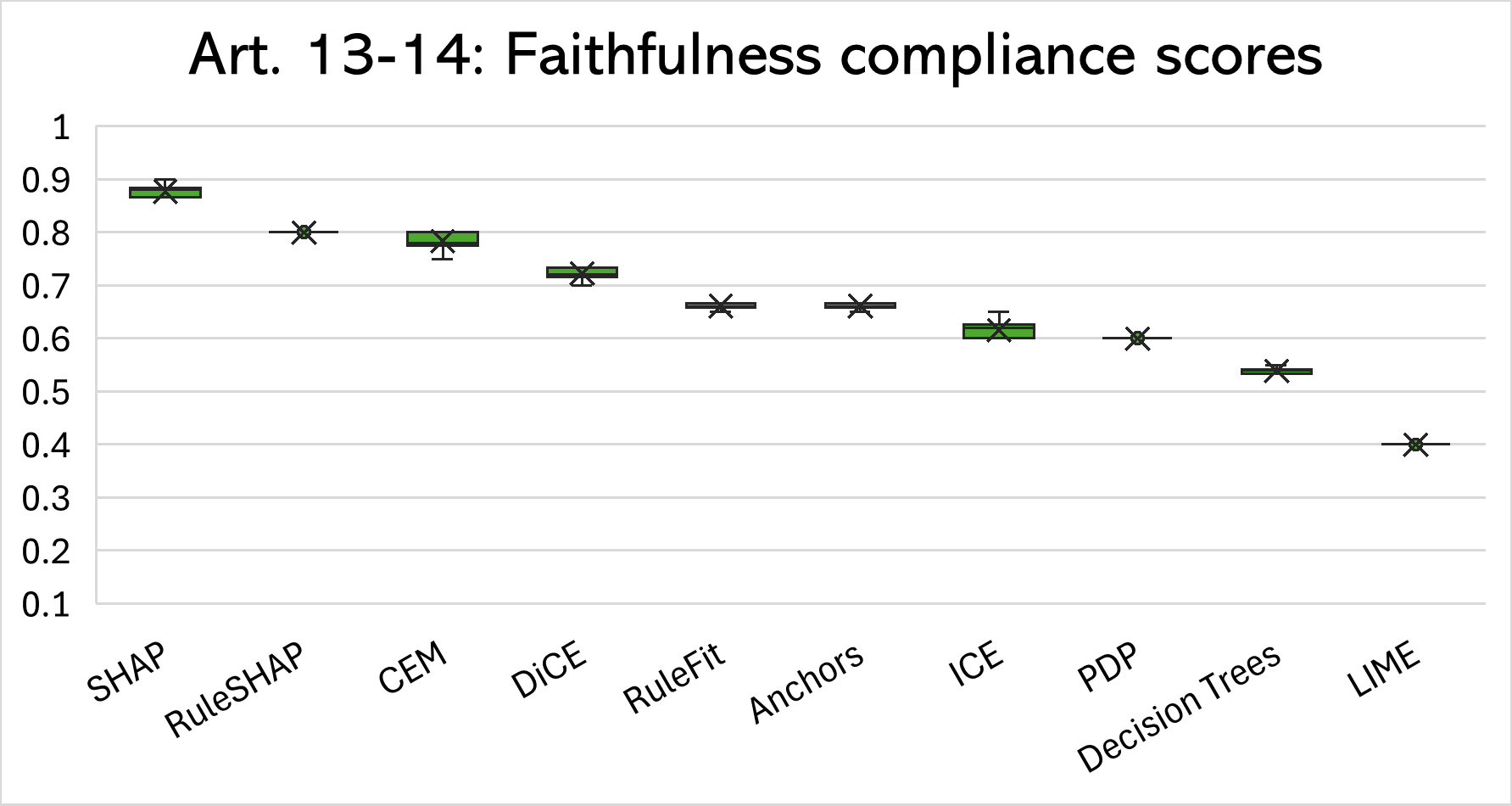}
    \caption{}
    \end{subfigure}
    \begin{subfigure}[b]{0.49\textwidth}
    \includegraphics[width=\linewidth]{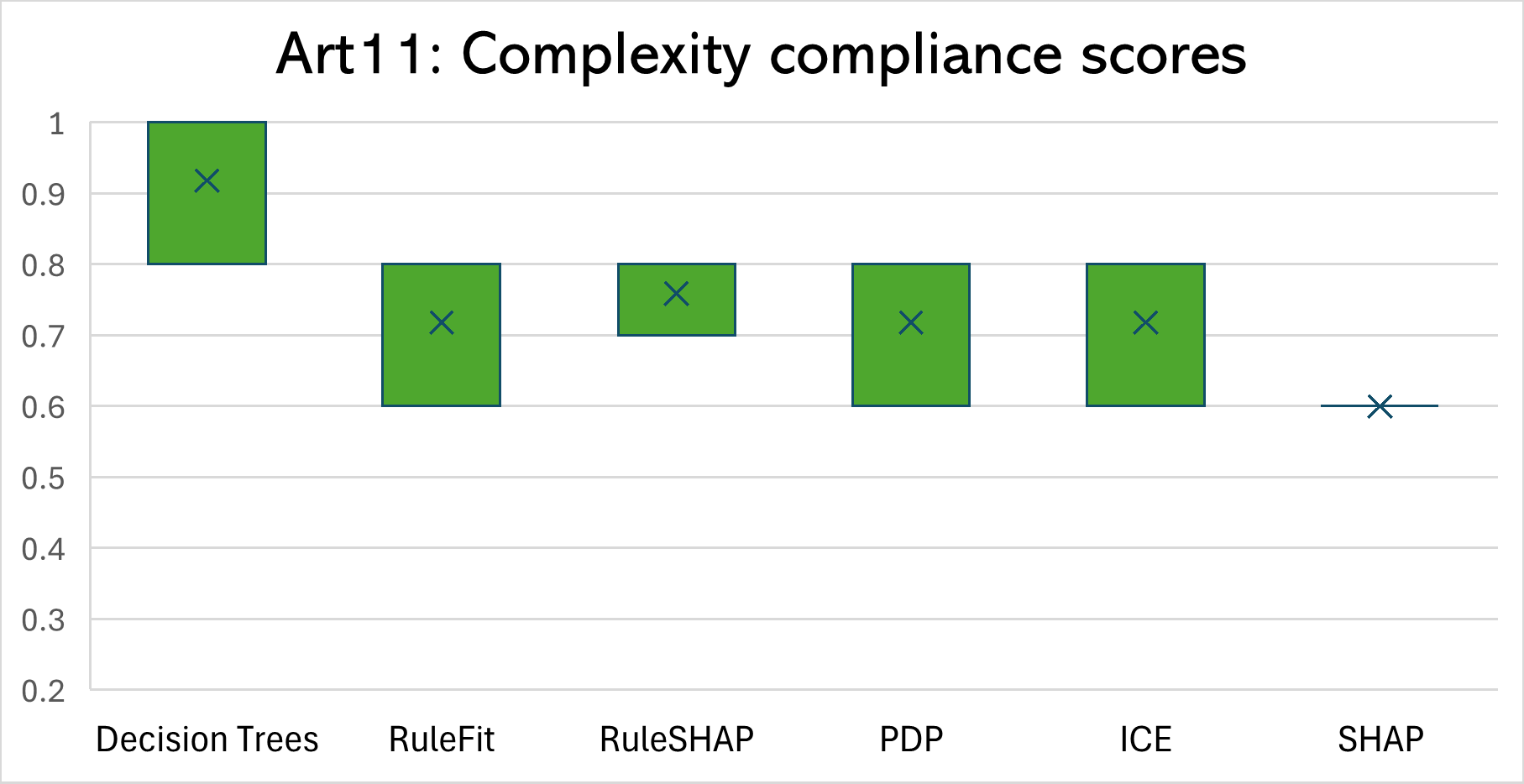}
    \caption{}
    \end{subfigure}
    \caption{Results of the sensitivity analysis over the compliance scores related to (a) faithfulness, (b) robustness and (c) complexity as demanded by Art.\ 86 of the AI Act, (d) faithfulness demanded by Arts.\ 13--14 and (e) complexity demanded by Art.\ 11.}
    \label{fig:delta_variation}
\end{figure}

The $\delta$ variations had some impact on the other five combinations, namely faithfulness, robustness and complexity for Art.\ 86, faithfulness as required by Arts.\ 13--14 and complexity relative to Art.\ 11. However, such an impact was insufficient to change the ranking of the XAI methods, as shown in Figure~\ref{fig:delta_variation}. SHAP, RuleSHAP and CEM are still the favourite XAI methods for meeting the faithfulness requirements of Art.\ 86 and Arts.\ 13--14 and SHAP, RuleFit and RuleSHAP best meet the robustness demands of Art.\ 86. Anchors and Decision Trees remain the two optimal XAI methods in terms of complexity as demanded by Arts.\ 11 and 86, despite the variations in the $\lambda$ factors.
Overall, the sensitivity analysis shows that the proposed compliance scores are robust to variations in the legal strength factors $\lambda$.
Although our 1--5 scores are ordinal and context-dependent, aggregating them into a quantitative index enables comparative analysis of XAI profiles against the AI Act's explanatory requirements.

\section{Conclusion and Future Work} \label{sec:conclusion}
We addressed the critical \emph{transparency gap} between the technical capabilities of state-of-the-art XAI methods and the diverse explainability requirements imposed by the AI Act. By conducting a systematic review of both XAI algorithms and their intrinsic properties, alongside a parallel analysis of the AI Act's relevant provisions, we derived a harmonised set of \emph{legal explanatory requirements}. Our interdisciplinary mapping then translated these requirements into measurable XAI algorithm properties (faithfulness, robustness, and complexity), enabling a principled alignment between legal mandates and technical artefacts.
Methodologically, this alignment is achieved through a mixed-methods design that combines qualitative legal and literature-based assessment with quantitative score aggregation.

The proposed law-to-XAI mapping pipeline is deliberately modular and extensible. New legal requirements or interpretability properties can be incorporated by updating the relevant rows or columns in Table~\ref{tab:property-mapping} and assigning the appropriate \checkmark{} and \xmark{} indicators. Weighting factors $\lambda_s$ may be derived from legal precedent (e.g., court decisions) or specified directly by legal experts.

New XAI algorithms can likewise be integrated through a one-time evaluation using the same assessment rubric, after which scores are automatically recomputed via Eq.~\eqref{eq:score}. Because the scoring depends only on the legal requirements $\mathcal{R}_r$, the property weights $w_{p_i}$, and the procedural fit indicator, no additional manual calibration is needed.
This design also supports rapid adaptation to legal change: when legislation evolves, only the corresponding weighting factors need to be updated, thus enabling immediate assessment of whether existing XAI tools remain sufficient or additional safeguards are required.

Future research avenues include testing the proposed compliance scoring framework in real-world AI deployment case studies, where 
legal and technical experts will assess whether the scores meaningfully track regulatory requirements and are usable in high-risk systems.



%
%
\bibliographystyle{splncs04nat}
\bibliography{bibliography}
\end{document}